# DDPG based on multi-scale strokes for financial time series trading strategy

JUN-CHENG CHEN

Faculty of Information Technology, Beijing University Of Technology

Email: juncheng@bjut.edu.cn

CONG-XIAO CHEN, LI-JUAN DUAN, ZHI CAI

Faculty of Information Technology, Beijing University Of Technology

Email: 1140202590@qq.com, ljduan@bjut.edu.cn, caiz@bjut.edu.cn

With the development of artificial intelligence, more and more financial practitioners apply deep reinforcement learning to financial trading strategies. However, It is difficult to extract accurate features due to the characteristics of considerable noise, highly non-stationary, and non-linearity of single-scale time series, which makes it hard to obtain high returns. In this paper, we extract a multi-scale feature matrix on multiple time scales of financial time series, according to the classic financial theory-Chan Theory, and put forward to an approach of multi-scale stroke deep deterministic policy gradient reinforcement learning model (MSSDDPG) to search for the optimal trading strategy. We carried out experiments on the datasets of the Dow Jones, S&P 500 of U.S. stocks, and China's CSI 300、SSE Composite, evaluate the performance of our approach compared with turtle trading strategy, Deep Q-learning (DQN) reinforcement learning strategy, and deep deterministic policy gradient (DDPG) reinforcement learning strategy. The result shows that our approach gets the best performance in China CSI 300、SSE Composite, and get an outstanding result in Dow Jones, S&P 500 of U.S.

**ACM Reference Format:**
Jun-cheng Chen, Cong-xiao Chen, Li-Juan Duan, Zhi Cai. Faculty of Information Technology,Beijing University Of Technology, The Title of the Paper: DDPG based on multi-scale strokes for financial time series trading strategy In ICCTA 2022: 2022 8th International Conference on Computer Technology and Applications, May 12-14, 2022, Vienna, Austria(Hybrid). ACM, New York, NY, USA, XX pages. https://doi.org/ (the doi information will be completed after uploading copyright to press).

## 1 INTRODUCTION

Time series is a set of random variables sorted by time. The analysis of time series is to use statistical means to analyze the past of this series to model the changing characteristics of this variable and predict the future. Time series analysis has a wide range of applications in many fields, among which the analysis of the stock market is a common and typical financial problem in the financial field. According to the Efficient Markets Hypothesis [1]: all valuable information has been timely, accurate, and fully reflected in the stock price trend, so the past state of the financial time series can be modeled to obtain stock trading strategies. A good stock trading strategy can control risks and maximize returns. Therefore, it is very important for the majority of financial practitioners to choose a good stock trading strategy.

Computer algorithms used in automatic trading have gradually become popular in recent years. There are many traditional trading strategies, such as moving average [2], relative strength index [3], turtle trading strategy

[4], etc., which generate trading signals based on financial indicators. With the development of artificial intelligence, more and more researchers and experts have applied machine learning algorithms to stock trading [5-10]. Reinforcement learning is good at solving nonlinear problems with delayed returns, and it can effectively improve the adaptability of financial transaction models by using existing information to obtain the best returns. Moody al. [11]proposed the algorithm of cyclic reinforcement learning. Eiler et al. [12]combined artificial neural networks with reinforcement learning (RL) ideas. The value iteration method of RL is used to train the artificial neural network, and only the immediate return is optimized. The back propagation algorithm is used to train the neural network. Wang et al. [13]built an automated trading system based on Deep Q-learning, and proved that the method based on deep Q-learning is better than buy-and-hold strategy and RRL strategy. Deng Y al. [14] introduced the contemporary DL into a typical DRL framework for financial signal processing and online trading, and validated the robustness of the neural system in future markets for stocks and commodities. Xiong Z al.[15]trained a deep reinforcement learning agent and obtained an adaptive trading strategy. YuanY al. [16]. proposed a framework called reinforcement learning based on data augmentation. And it proved that Deep Q-learning and soft actor critic (SAC) can beat the market in Sharp Ratio. Singhania R al.[17] forecasted time serirs by using recurrent neural networks with google trends data. Li Y al. [18] implemented a novel deep reinforcement learning for stock transaction strategy, proved the practicality of DRL in dealing with financial strategy issues, and compared three classical DRL models. Hyungjun Park al. [19] proposed a new approach for deriving a multi-asset portfolio trading strategy using deep Q-learning.

However, most of the above studies are based on single-scale processing, which will lead to losing multi-scale information and increase the difficulty of reinforcement learning. For example, the K-Line is in an upward trend on the daily line, maybe in a downward trend on the weekly line. It means that it is in a short-term rise and long-term decline trend. Selling stock may be the correct response due to the future decline expectations. The model based on single-scale usually focuses on the short-term trend and ignores the long-term trend, which may send buying signal before the decline or send a selling signal before upward. The characteristic of the model makes it hard to get high returns. In order to overcome the shortcoming of the single-scale model, we propose an approach based on multi-scale time series. In our approach, we extract strokes from multi-scale time series according to Chan Theory, combine the strokes of all time scales into a feature matrix as the state variable of Markov for reinforcement learning.

In this paper, we first introduce the pre-processing technique of K-Line data in Chan Theory, including the relationship among adjacent K-lines, bottom-shape, top-shape, and constructing technique of a stroke. Then, we model stock trading as a Markov decision process and take the feature matrix, which includes original K-Line data and multi-scale strokes mentioned above, as the input parameter. Subsequently, we introduce the algorithm of DDPG in detail and put forward our trading strategies based on DDPG algorithm [21]. Finally, we carry out experiments on the specific financial time series of CSI300, SSE Composite, DJI, and S&P 500, and compare it with traditional turtle trading strategy, DQN deep reinforcement learning strategy, and the buy-and-hold strategy.

## 2 MULTI-SCALE FEATURE EXTRACTION MODULE

In our approach, the stroke is the feature matrix's key element and is extracted based on K-Line inclusion, bottom-shape, and top-shape. So the following of this section first introduces K-Line inclusion, bottom-shape, and top-shape, then describes the concept of the stroke, and depicts multi-scale processing technology at last.

## 2.1 Inclusion relationships

After the financial time series is transformed into K-Line series, it can be found that the stock market volatility has the nature of random and unstable, and has a lot of useless K-Line data which affects the trend judgment of the long-term stock price. The K-Line series with complex fluctuations can be processed with standardized method, which can be more conducive to find long-term trends and standardized classification. This paper uses the idea of methods in Chan Theory which is to remove the inclusion relationships of financial time series.

We define $[ki_l, ki_h]$ as the lowest value, highest value of the i-th K-Line. With the symbols defined above, inclusion relationship of continuous K-Line is defined as:

$$ki_h < k(i+1)_h \text{ and } ki_l > k(i+1)_l$$
$$ki_h > k(i+1)_h \text{ and } ki_l < k(i+1)_l$$

The method of removing the inclusion relationship needs to be discussed separately according to the two trends of the K-Line. If it occurs in the ascending relationship sequence, the method of removing the inclusion relationship and forming a new K-Line (named k') is calculated as:

$$k'_l = \max(ki_l, k(i+1)_l) \text{ and } k'_h = \max(ki_h, k(i+1)_h)$$

If it occurs in the descending relationship sequence, the calculation of the new bar that removes the inclusion relationship is (Figure 1):

$$k'_l = \min(ki_l, k(i+1)_l) \text{ and } k'_h = \min(ki_h, k(i+1)_h)$$

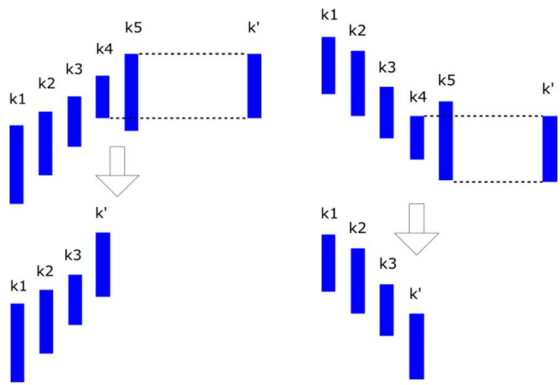
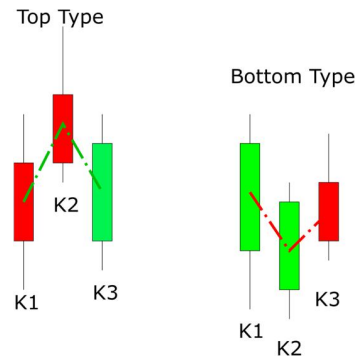

Figure 1: Remove inclusion relationship       Figure 2: Top-bottom shape

## 2.2 Bottom-shape and top-shape

In a K-Line sequence with no inclusion relationship, we can abstract two kinds of shapes from it. Top-shape is defined as (Figure2):

$$ki_h < k(i+1)_h \text{ and } k(i+2)_h < k(i+1)_h \text{ and } ki_l < k(i+1)_l \text{ and } k(i+2)_l < k(i+1)_l$$

Bottom-shape of three consecutive bars is defined as:

$$ki_h > k(i+1)_h \text{ and } k(i+2)_h > k(i+1)_h \text{ and } ki_l > k(i+1)_l \text{ and } k(i+2)_l > k(i+1)_l$$

The top-bottom-shape is the turning point, which is the market's psychological expectations of the current K-Line trend.

**2.3 Extract strokes**

To mark trend information, we need to connect adjacent turning points as strokes, but to meet the strict strokes, the requirements include three parts:

(1) There must be alternating top-shape and bottom-shape between two adjacent shapes. Therefore, for top-shape, if there is a high point of top-shape behind, the stroke is directly connected to the higher top-shape. If there is a lower bottom behind the Bottom point of the bottom-shape, the stroke is directly connected to the lower bottom-shape. The stroke segment from top-shape to bottom-shape is called a descending stroke. On the contrary, the stroke segment from bottom-shape to top-shape becomes a rising stroke.

(2) After processing the inclusion relationship, there must be at least one non-public K-Line between the top-bottom-shape. The specific method is to have at least five K-Line from the lowest K-Line of the bottom-shape to the highest K-Line of the top-shape. Otherwise, one stroke segment cannot be completed, and the stroke should be connected to the next shape.

(3) The adjacent top-shape and bottom-shape shall ensure that the high price of the top-shape is higher than the low price of the bottom-shape.

Figure 3 is a diagram of drawing a strict stroke for part of the data of CSI 300. stroke splitting describes the important turning point information of the original data in the form of a piecewise linear function for the original stock price series. These turning points information replaces the original redundant data information, making the time series prediction clearer.

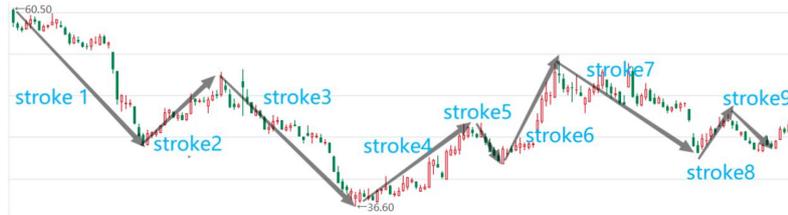

Figure 3:Strokes

**2.4 Multiscale processing technique**

In order to obtain the macro information of financial time series, it is necessary to extract multi-scale strokes on financial time series .As shown in Figure 4, We show the multi-scale strokes extraction results of three different scales (10 minutes scale, day scale , week scale), then the total characteristic matrix is $O = \{o^1, o^2, o^3 ... o^n\}$, $o^1$ represents the original 10-minute scale K-Line data, $o^2$ represents the stroke data on the 10-minute scale, which contains open price、close price、high price、low price、volume and Stroke trend judgment, $o^3$ represents Stroke data on the day-level scale, $o^4$ represents Stroke data on the week-level scale. Each level of feature has a different meaning. $o^4$ contains the long-term trend of financial time series, $o^3$ contains the mid-term direction characteristics of financial time series, low-scale $o^2$, $o^1$ guides short-term trading operations. In this module, Single-scale data is expanded to multi-scale data, and the Strokes extracted at each scale are combined into a feature matrix for output, effectively extracting the long-term and short-term trend characteristics of financial time series, eliminating unnecessary complex noise.

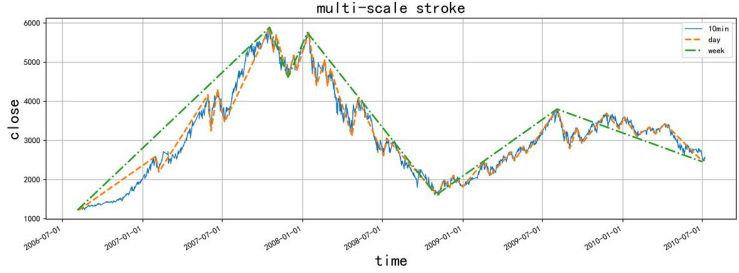

Figure 4: Multi-scale strokes

## 3 DEEP DETERMINISTIC POLICY GRADIENT BASED ON MULTI-SCALE STROKE

In this section, the proposed multi-scale stroke deep deterministic policy gradient reinforcement learning model(MSSDDPG) is demonstrated. Based on the multi-scale feature matrix mentioned above and the characteristics of the market environment, we model the stock trading problem as a Markov Decision Process (MDP) to solve the RL problem, in which MDP is defined as $(S, A, R, \pi, Q(\pi))$, where:

(1) State space $S = [O, h, M]$ : $O = \{o^1, o^2, o^3 \dots o^n\}$ :The output of the multiscale fused feature module described in the previous chapter, where n represents the number of levels divided into multiscale levels, $o^1$ represents the initial K-line data of the stock at the current time t. $o^2 \sim o^n$ is the data of strokes at different levels at the current time t, including the opening price, closing price, highest price, lowest price, volume and volume of Strokes at the current scale. $h \in Z_+$ is the number of stocks held and $M \in Z_+$ is the remaining capital.

(2) Action space $A = [\alpha, W]$: $\alpha$ defines a series of operations on stocks, which represent buying, holding and selling respectively, and $W$ represents the trade share.

(3) Rate of return $(s, \alpha, s^{'})$ : the rate of return generated by taking action $\alpha$ in state $s$ and entering state $s^{'}$.

(4) Trading strategy $\pi(s)$: represents the mapping from the state space $s$ to the action space $\alpha$.

(5) State action value function $Q_\pi(s, \alpha)$: the expected reward for executing action $\alpha$ in the state $s$ according to the strategy.

This paper chose the deep deterministic strategy gradient algorithm [21] as the strategy algorithm of reinforcement learning. The DDPG algorithm is closely related to Q learning [22] and policy gradient [23]. The algorithm framework is mainly composed of two parts: Actor network and Critic network. In the strategy gradient algorithm, a random strategy is used to output the probability of action. At each step, the best action in the current state is obtained according to the probability distribution. The random strategy is adopted to generate the action:

$$\alpha_t \sim \pi_\theta(s_t | \theta^\pi)$$

DDPG adopts a Deterministic Policy Gradient, that is, the behavior of each step directly obtains a certain value through the optimal behavior strategy:

$$\alpha_t = \mu(s_t | \theta^\mu)$$

Where $\mu$ is a function representing the optimal behavior strategy. In DDPG, a convolutional neural network $\mu$ is used to simulate the function, and its parameter is $\theta^\mu$.

The objective function for DDPG is:

$$J(\mu_\theta) = \int_S \rho^\mu(s) Q^\mu(s, \mu_\theta(s)) ds = E_{S \sim \rho^\mu}[Q^\mu(s, \mu_\theta(s))]$$

Where $\rho^\mu(s)$ refers to the distribution function of the environment s generated under the strategy $\mu$. To determine the optimal behavior strategy $\mu$ is to maximize $J_\theta(\mu)$:

$$\mu = argmax J_\theta(\mu)$$

Therefore, the deterministic action strategy gradient is:

$$\nabla_\theta J(\mu_\theta) = \int_S \rho^\mu(s) \nabla\theta \mu_\theta(s) Q^\mu(s,\alpha)|_{a=\mu\theta} ds = E_{s\sim\rho^\mu}[\nabla\theta\mu_\theta(s)Q^\mu(s,\alpha)|_{\alpha=\mu\theta}]$$

In the DDPG algorithm, the above-mentioned network for determining the optimal behavior strategy is regarded as the Actor-network, and DDPG adopts the same structure as DQN and uses a convolutional neural network to simulate the Q function, which is called a deep Q network. DDPG uses this network to simulate Combining the above function, that is, fitting the expected value function of the reward for the selection action under the deterministic strategy μ, so the value function in the DDPG network acts as a critic network.

## 4　DATA AND RESULT

The buy-and-hold strategy is to buy and hold the asset allocation of the investment portfolio for a long time, keeping the same direction and the same proportion of changes in the stock market value, giving up the possibility of profiting from changes in the market environment. We use this strategy to express the net value of the index and show the change in the value of the stock market.

We use the traditional financial trading strategy-Turtle trading strategy, DQN trading strategy, DDPG trading strategy as baseline trading strategies. The Turtle trading strategy is a well-known public trading system, and its trading rules are:

(1) if today's close price is greater than the highest price in the past 20 trading days, buy at the close price;

(2) if the close price is lower than the lowest price in the past 10 trading days, sell at the close price.

The turtle trading strategy combines batch opening, dynamic stop-profit and stop-loss, and trend follow-up of the market. It is a complete and effective trading system.

DQN and DDPG have shown great power in many tasks, and many researchers have applied them to scenarios related to trading algorithms. In our experiments, we selected the k-Line data of the past 30 days as the feature for training.

We tested these four trading strategies on four test data sets and compared in the net value of the index, cumulative return, annual return, maximum drawdown rate, Sharpe ratio[24], Alpha, and Beta.

### 4.1 Experimental data

We tested the MSSDDPG trading model on real world financial data -The S&P 500 index、Dow Jones Industrial Index On US stocks and CSI300 index(sh000300)、SSE Composite(sh000001)in Chinese stock market. The above data sets are from the AKShare, an open-source financial database. To verify the performance of the model under the long-term trend and ensure the credibility of the results, the data set we selected has a long time span. The average time span of the training set of the four data sets is 16 years. The average time span of each test set is seven years, and each test set spans a bull market and a bear market, which perfectly simulates the volatility of stocks (Table 1).

Table 1:Data set

| Data set | Train set | Test set |
| --- | --- | --- |
| CSI300 | 2005.01.01-2014.01.01 | 2014.01.01-2021.12.01 |
| SSE Composite | 1990.01.01-2015.01.01 | 2015.01.01-2021.12.01 |
| DJI | 1990.01.01-2014.01.01 | 2014.01.01-2021.12.01 |
| S&P 500 | 2001.01.01-2008.01.01 | 2008.01.01-2014.01.01 |

## 4.2 Experimental results and analysis

Annualized return, max drawdown, Sharpe ratio, Alpha, and Beta are used as evaluation metrics. Annualized return represents the mean return each year during the long time. Max drawdown indicates maximum pullback of the net value. Sharpe ratio is a kind of risk-adjusted return, it is applied to compare risk-adjusted returns of different trading strategies. Alpha is used in finance as a measure of performance, indicating when a strategy, trader, or portfolio manager has managed to beat the market return over some period. It is often considered the active return on an investment, gauges the performance of an investment against a market index or benchmark that is considered to represent the market's movement as a whole. A security's Beta is calculated by dividing the product of the covariance of the security's returns and the market's returns by the variance of the market's returns over a specified period. It effectively describes the activity of a security's returns as it responds to swings in the market.

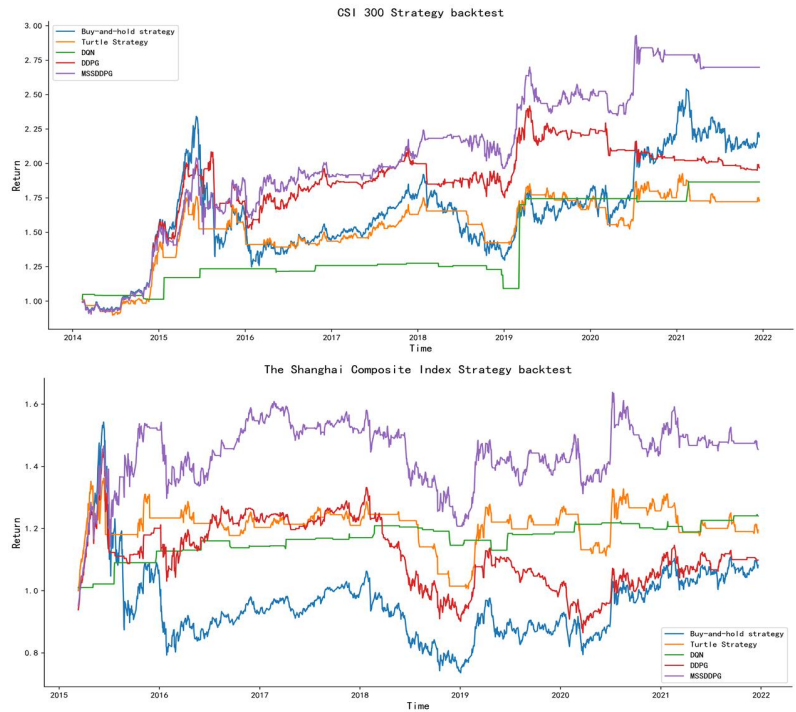

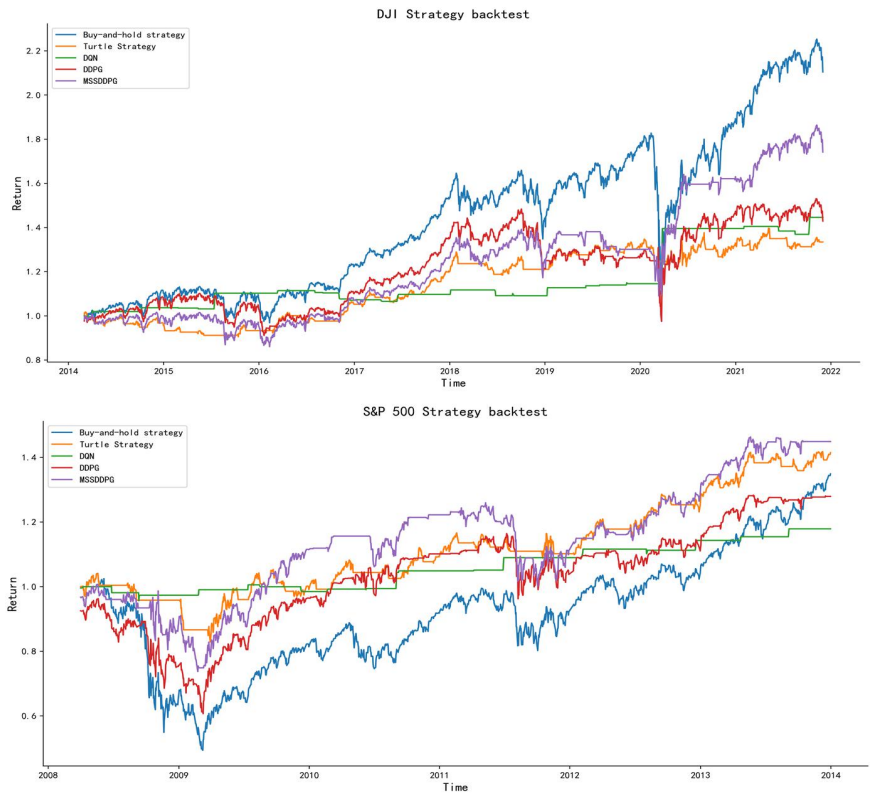

Figure 5 : From top to bottom are CSI300, The Shanghai Composite Index, DJI and S&P 500

Table 2 : Backtest

| Stock | Strategy | Cumulative return | Annual return | Max drawdown | Alpha | Beta | Sharpe |
|---|---|---|---|---|---|---|---|
| CSI 300 (2014.01.01-2021.12.01) | B&H | 120.28% | 10.85% | 46.7% | - | - | - |
| | Turtle | 73.53% | 7.46% | 21.56% | 0.03 | 0.38 | 1.01 |
| | DQN | 86.53% | 8.47% | 14.36% | 0.08 | 0.0 | 1.25 |
| | DDPG | 96.83% | 9.24% | 26.93% | 0.035 | 0.528 | 1.44 |
| | MSSDDPG | 169.85% | 13.83% | 27.2% | 0.07 | 0.61 | 1.87 |
| The Shanghai Composite Index (2015.01.01-2021.12.01) | B&H | 8.23% | 1.2% | 52.3% | - | - | - |
| | turtle | 19.59% | 2.74% | 26.1% | 0.024 | 0.323 | 0.1 |
| | DQN | 24.04% | 3.31% | 6.53% | 0.033 | 0.002 | 0.12 |
| | DDPG | 9.84% | 1.43% | 40.23% | 0.007 | 0.62 | 0.12 |
| | MSSDDPG | 45.44% | 5.83% | 24.95% | 0.051 | 0.641 | 0.74 |

| Market | Strategy | | | | | |
|---|---|---|---|---|---|---|
| DJI (2014.01.01-2021.12.01) | B&H | 110.43% | 9.99% | 37.09% | - | - | - |
| | Turtle | 33.45% | 3.76% | 11.05% | 0.015 | 0.223 | 0.35 |
| | DQN | 44.59% | 4.83% | 4.39% | 0.047 | 0.016 | 0.85 |
| | DDPG | 42.9% | 4.67% | 34.3% | -0.03 | 0.768 | 0.55 |
| | MSSDDPG | 74.17 | 7.36 | 21.67 | 0.003 | 0.764 | 1.0 |
| S&P 500 (2008.01.01-2014.01.01) | B&H | 34.9% | 5.35% | 52.58% | - | - | - |
| | Turtle | 41.45% | 6.22% | 20.37% | 0.051 | 0.207 | 0.79 |
| | DQN | 17.89% | 2.91% | 4.61% | 0.029 | 0.001 | -0.01 |
| | DDPG | 27.92% | 5.76% | 33.48% | 0.047 | 0.604 | 0.56 |
| | MSSDDPG | 44.91% | 6.67% | 26.79% | 0.035 | 0.597 | 0.75 |

Figure 5 and Table 2 show that:

1) MSSDDPG has achieved good performance in Chinese market(The Shanghai Composite Index, CSI 300). Its cumulative return rate, annualized return rate, and other financial evaluation indicators show that MSSDDPG can take the initiative to obtain high returns and avoid risks in time when the index retracements.
2) MSSDDPG surpasses other strategies but does not exceed market performance.
3) MSSDDPG performed slightly lower than the Turtle Strategy in Sharpe ratio, but get higher annual return. on the S&P 500.
4) MSSDDPG performed best in the Chinese market, but slightly inferior in the US stock market.

The results indicate that our approach get the highest annual return among the four strategies. It seems that:

1) Our strategy is better at seeking returns in volatile markets. In rising markets, it may miss some opportunity of transaction, in order to control its risk. Generally speaking, MSSDDPG is very adaptive to risk. Its maximum drawdown in all markets is less than the index net value, and the Sharpe ratio is greater than 70%, which is suitable for obtaining stable returns in volatile markets.
2) DQN is the most stable strategy, with the lowest volatility, and is less affected by the index retracement, but the rate of return is also less. It tends to gain profits with a small number of operations.
3) As a traditional financial trading strategy, the turtle strategy is relatively balanced, but it cannot beat the MSSDDPG strategy in most cases. It only achieves almost the same performance on the S&P 500, which shows that deep reinforcement learning can indeed defeat the traditional financial investment strategy.

## 5 CONCLUSION AND FUTURE

In this paper, we briefly discuss the research methods of financial time series and illustrate the limitations of a single-scale model, then we use the multi-scale feature extract module to combine the deep deterministic strategy gradient algorithm for trading strategies of financial time series, which proves the effectiveness, superiority, and practicability of the MSSDDPG model in different market conditions. Compared with other studies, we pay more attention to long-term effects, and the test set has a longer time span, hoping to better reflect the performance of the trading strategy model under the long-term trend. The conclusion is a good demonstration of the feasibility of artificial intelligence in the direction of quantitative finance.

In the future, we will carry out mixed forecasting exploration on multiple stocks, complete the combination of stock selection; and find a more perfect trading strategy in the upward trend, and obtain results that exceed market performance; we can also make a more explanatory multi-scale model, find more accurate buying and selling points, to increase the practicality of the model.